\documentstyle[sprocl,psfig]{article}

\def\be{\begin{equation}}
\def\ee{\end{equation}}
\def\bea{\begin{eqnarray}}
\def\eea{\end{eqnarray}}

\def\gtrsim{\mathrel{\hbox{\rlap{\hbox{\lower4pt\hbox{$\sim$}}}\hbox{$>$}}}}

\begin{document}
\title{NO EVOLUTION IN THE X-RAY CLUSTER LUMINOSITY FUNCTION OUT TO $z\simeq0.7$}
\author{ROMER, A.K., NICHOL R.C.,}
\address{Department of Physics, Carnegie Mellon University, 
5000 Forbes Avenue, Pittsburgh, PA 15213, USA}
\author{COLLINS, C.A., BURKE, D.J.,}
\address{Astrophysics Group, School of Electrical Engineering, Electronics 
and Physics, Liverpool John Moores University, Byrom Street, Liverpool, L3 3AF, UK}
\author{HOLDEN, B.P.,} 
\address{Department of Astronomy and Astrophysics, University of Chicago, 5640 S. Ellis Avenue, Chicago, IL 60637, USA}
\author{METEVIER, A., ULMER, M.P., PILDIS, R.}
\address{Department of Physics and Astronomy, Northwestern University, 
2145 Sheridan Road, Evanston, IL 60208, USA}
 

\maketitle\abstracts{In contrast to claims  of
strong negative evolution by other groups,
we find no evidence for evolution of the X-ray cluster luminosity 
function out to $z\simeq0.7$.}

The Serendipitous High-redshift Archival ROSAT Cluster (SHARC) survey
 (http://www.astro.nwu.edu/sharc)
is a project to identify $\gtrsim 100$ X-ray clusters at redshifts
greater than $z=0.3$.  Although similar in aims and approach, it is 
much larger than the RDCS\,\cite{pr95}, WARPS\,\cite{cas97} and
RIXOS\,\cite{fjc96} surveys: when complete it will 
cover $\gtrsim 200$ square degrees. 
The SHARC survey X-ray data pipeline is fully automated and
is based on the EXAS\,\cite{sls94} reduction package and a wavelet 
transform source detection algorithm. The pipeline has been thoroughly 
tested\,\cite{rcn97} and has already been applied\,\cite{akr97} to 530 high 
galactic latitude PSPC pointings. A further $\simeq500$ PSPC and $\simeq200$ 
HRI pointings will be added to the survey in 1997. Optical follow-up is 
underway at the ARC 3.5m, ESO 3.6m and AAT 3.9m telescopes.



\begin{figure} 
\protect{
\centerline{\psfig{figure=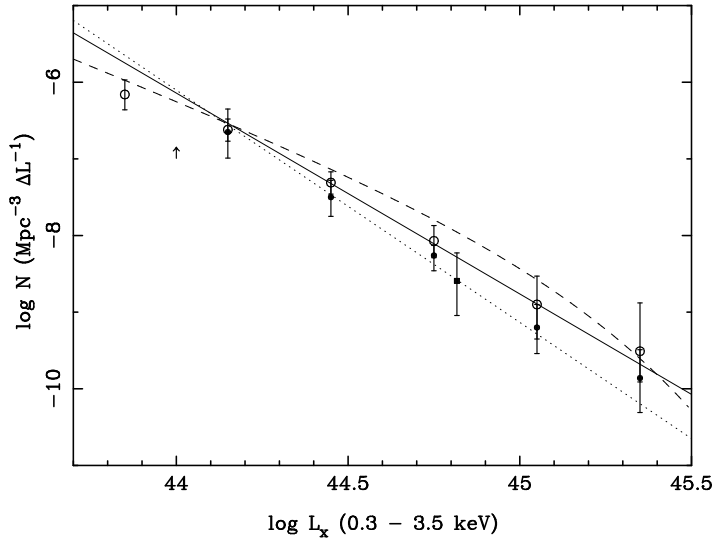,height=2.78in,angle=270.}}}
\caption{High redshift XCLF (solid line and symbols) and low redshift
XCLF (dotted lines, open symbols) for the EMSS cluster sample. Also
shown is the low redshift XCLF$^{6}$ for RASS clusters (dashed line).
(See Nichol {\it et al.} 1997 for more details.)}
\end{figure}


The primary research goal of the SHARC survey is the robust quantification
of X-ray cluster evolution. Reports of rapid negative evolution 
in the  X-ray cluster luminosity function (XCLF) seen in 
the EMSS\,\cite{img90}~\cite{pjh92} and RIXOS\,\cite{fjc96} cluster samples 
have attracted much attention because they are not consistent with standard 
hierarchical models of structure formation\,\cite{nk86}. Unfortunately,
both the EMSS and RIXOS samples suffer from small number statistics.
With the availability of thousands of deep pointings in the ROSAT archive,
the subject is ripe for re-examination.
Nichol {\it et al.} (1997) discusses a re-analysis of the properties of 
the EMSS cluster sample. 
A large fraction of the EMSS clusters were observed 
during ROSAT pointed observations. By applying the SHARC data 
analysis techniques to these observations, we were able 
to determine the cluster fluxes more accurately than was possible with 
the original Einstein data. Using these revised fluxes, and incorporating
new optical data where available, we produced the XCLFs shown in
Figure 1. As can be seen from that figure, the EMSS low and high redshift 
XCLFs are very similar, indeed they differ by only $1\sigma$.
A comparison between the EMSS high redshift XCLF and a local 
XCLF\,\cite{he97} based on ROSAT All-Sky Survey (RASS) data yields no
significant evidence for evolution out to $z\simeq0.5$. 

Collins {\it et al.} (1997) describes the results of a redshift survey\,\cite{djb96} of extended sources found in 66 southern 
PSPC pointings. The  survey yielded a sample of 36 clusters over 17 
square degrees to a flux limit of $\simeq3.9\times10^{-14} {\rm ergs\,sec^{-1}\,cm^{-2}}$.
Sixteen of the 36 clusters lie in the redshift range $0.30<z<0.67$. 
This is in stark contrast to the RIXOS survey\,\cite{fjc96}, which  
found only four $z\geq0.3$ clusters over a similar area (see Figure 2). 
Using predictions based on extrapolation of two local\,\cite{dg96}~\cite{he96} XCLFs, we conclude that our sample shows no evidence for 
significant evolution out to $z\simeq0.7$.

The studies described above challenge previous claims
of a rapidly evolving XCLF. They provide momentum to
our continuing ambition to provide the public with the
largest distant X-ray cluster sample possible. This
sample will have a well understood selection function and
will be vital not only to evolution studies but also
to studies of superclustering, 
the Butcher-Oemler effect, the cluster temperature function and 
the Sunyaev-Zeldovich effect.\\

\centerline{
This work was partially supported by NASA ADP grant NAG5-2432.\\}

\begin{figure} 
\protect{
\centerline{\psfig{figure=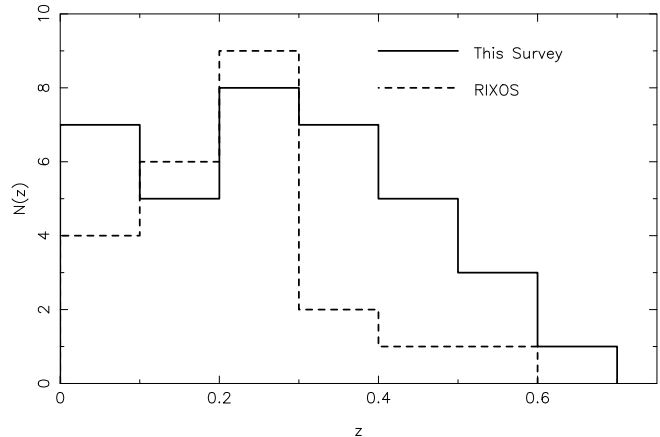,height=2.25in,angle=270.}}}
\caption{Comparison of the redshift histograms for southern SHARC$^{1}$
(solid line) and RIXOS$^{2}$
(dotted line) cluster samples. (See Collins {\it et al.} 1997 for details.)}
\end{figure}


\end{document}